\documentclass[11pt,a4paper,fleqn]{article}
\input epsf                                
\catcode`\@=11
\@addtoreset{equation}{section}
\def\theequation{\arabic{section}.\arabic{equation}}
\def\appendix{\renewcommand{\thesection}{\Alph{section}}\setcounter{section}{0}
              \renewcommand{\theequation}
            {\mbox{\Alph{section}.\arabic{equation}}}\setcounter{equation}{0}}

\def\maketitle{\thispagestyle{empty}\setcounter{page}0\newpage
                \renewcommand{\thefootnote}{\arabic{footnote}}
                  \setcounter{footnote}0}
\renewcommand{\thanks}[1]{\renewcommand{\thefootnote}{\fnsymbol{footnote}}
               \footnote{#1}\renewcommand{\thefootnote}{\arabic{footnote}}}

\renewcommand{\title}[1]{\begin{center}\Large\bf #1\end{center}\rm\par\bigskip}
\renewcommand{\author}[1]{\begin{center}\Large #1\end{center}}
\newcommand{\address}[1]{\begin{center}\large #1\end{center}}

\def\babs{\hrule\par\begin{description}\item{Abstract: }\it}
\def\eabs{\par\end{description}\hrule\par\medskip\rm}
\renewcommand{\date}[1]{\par\bigskip\par\sl\hfill #1\par\medskip\par\rm}
\newcommand{\ack}[1]{\par\section*{Acknowledgments} #1}

\def\dinfn{Dipartimento di Fisica, Universit\`a di Trento\\
                           and Istituto Nazionale di Fisica Nucleare,\\
                                   Gruppo Collegato di Trento, Italia \medskip}

\def\csic{Consejo Superior de Investigaciones Cient\'{\i}ficas \\
Instituto de Ciencias del Espacio (ICE/CSIC)\\
 Campus UAB, Facultat Ciencies, Torre C5-Parell-2a planta, \\
08193 Bellaterra (Barcelona), Spain \medskip}

\def\ieec{Institut d'Estudis Espacials de Catalunya (IEEC), \\
Edifici Nexus, Gran Capit\`a 2-4, 08034 Barcelona, Spain \medskip}

\def\guido{Guido Cognola\thanks{e-mail: \sl cognola@science.unitn.it\rm}}
\def\sergio{Sergio Zerbini\thanks{e-mail: \sl zerbini@science.unitn.it\rm}}

\def\emilio{Emilio Elizalde\thanks{e-mail: \sl elizalde@ieec.uab.es\rm}}

\newcommand{\s}[1]{\section{#1}}

\def\M{{\cal M}}                       
\def\segue{\qquad\Longrightarrow\qquad} 
\def\hs{\qquad}               
\def\nn{\nonumber}            
\def\beq{\begin{eqnarray}}    
\def\eeq{\end{eqnarray}}      
\def\ap{\left.}               
\def\at{\left(}               
\def\aq{\left[}               
\def\cp{\right.}              
\def\ct{\right)}              
\def\cq{\right]}              
\def\R{{\hbox{{\rm I}\kern-.2em\hbox{\rm R}}}}   
\def\H{{\hbox{{\rm I}\kern-.2em\hbox{\rm H}}}}   
\def\N{{\hbox{{\rm I}\kern-.2em\hbox{\rm N}}}}   
\def\C{{\ \hbox{{\rm I}\kern-.6em\hbox{\bf C}}}} 
\def\Z{{\hbox{{\rm Z}\kern-.4em\hbox{\rm Z}}}}   
\def\ii{\infty}                                  
\def\X{\times\,}                                 
\def\Tr{\mathop{\rm Tr}\nolimits}                  
\renewcommand{\Im}{\mathop{\rm Im}\nolimits}       
\def\dir{/\kern-.7em D\,}                          
\def\lap{\Delta\,}                                 

\def\ze{\zeta}

\def\la{\lambda}

\def\Ga{\Gamma}

\def\La{\Lambda}

\begin{document}

\title{Multi-(super)graviton theory \\
on topologically  non-trivial backgrounds}

\author{\guido$^{,a}$, \emilio$^{,b,c}$ and \sergio$^{,a}$}
\address{\small $^a$\dinfn \\ $^b$\csic \\ $^c$  \ieec}
\babs
It is shown that in some multi-supergraviton models,
the contributions to the effective potential due to a non-trivial
topology can be positive, giving rise in this way to a
positive cosmological constant, as demanded by cosmological
observations.
\eabs

\s{Introduction}
Renewed interest in the study of
multi-graviton theories \cite{BDGH} owes, in particular,
 to the fact that
these formulations resemble higher-dimensional gravities in the presence of
 discrete dimensions. These classes of discretized Kaluza-Klein theories are
now in fact under the focus of attention due to their primary importance
for the realization of the dimensional deconstruction program
\cite{akio,AH}. Moreover, multigravitons can be also related with discretized
brane-world models \cite{DKP}.

In spite of the absence of a consistent interaction among the
gravitons, one can think of possible couplings in the theory space.
In particular, in a recent paper \cite{KS}, a multi-graviton theory
with nearest-neighbor couplings in the theory space has been
proposed. As a result, a discrete mass spectrum appears. The theory
seems to be equivalent to Kaluza-Klein gravity with a discretized
dimension.

In a previous paper concerning multi-graviton theory \cite{multig},
we have shown by means of an explicit example, namely a discretized
Randall-Sundrum (RS) brane-world \cite{RS1}, that the induced cosmological
constant becomes positive provided the number of massive gravitons
is sufficiently large.

In the present paper, we would like to show that an alternative mechanisms
can also give rise to positive contributions to the cosmological constant.
In particular we shall consider a multi-supergraviton example
with few gravitons, in a  manifold (bulk) with non trivial topology.
We shall show that in such a model
 a positive cosmological constant $\La$ can be generated,
due to the presence of
positive topological contributions. Moreover, by a suitable choice of
the topological parameters, the number obtained for $\La$ can reach a
value perfectly in accordance with result obtained from recent cosmological
observations  \cite{sdss1}.

\s{The multi-graviton and multi-supergraviton models}\label{S:models}

\subsection{The graviton model}

We start by considering the Lagrangian for the spin-two field $h_{\mu\nu}$
 with mass $m$
\beq
\label{KS8}
{\cal L}_m &=& {\cal L}_0 - {m^2 \over 2}\left(h_{\mu\nu}h^{\mu\nu} - h^2\right)
 - 2 \left(m A^\mu + \partial^\mu \varphi\right)
\left(\partial^\nu h_{\mu\nu} - \partial_\mu h\right) \nn\\ && -
{1\over2}\left(\partial_\mu A_\nu - \partial_\nu A_\mu\right)
\left(\partial^\mu A^\nu - \partial^\nu A^\mu\right) \,, \eeq where
${\cal L}_0$ is the Lagrangian of the massless spin-two field
(graviton) $h_{\mu\nu}$ ($h\equiv h^\mu_{\ \mu}$) \beq \label{KS7}
{\cal L}_0=-{1 \over 2}\partial_\lambda h_{\mu\nu}\partial^\lambda
h^{\mu\nu} + \partial_\lambda h^\lambda_{\ \mu}\partial_\nu
h^{\mu\nu}
 - \partial_\mu h^{\mu\nu}\partial_\nu h + {1 \over 2}\partial_\lambda h
\partial^\lambda h\ ,
\eeq
while $A_\mu$ and $\varphi$ are St{\"u}ckelberg fields \cite{hamamoto}.

The multi-graviton model is defined by taking $N-$copies of
(\ref{KS8}) with graviton $h_{n\mu\nu}$ and St{\"u}ckelberg fields
$A_{n\mu}$ and $\varphi_n$. Here, we propose a theory defined by a
Lagrangian which is taken to be a generalization of the one in
\cite{KS}. It reads \beq \label{KS9} {\cal L}&=&
\sum_{n=0}^{N-1}\left[ -{1 \over 2}\partial_\lambda h_{n\mu\nu}
\partial^\lambda h_n^{\mu\nu}
+ \partial_\lambda h^\lambda_{n\mu}\partial_\nu h_n^{\mu\nu}
 - \partial_\mu h_n^{\mu\nu}\partial_\nu h_n
+ {1 \over 2}\partial_\lambda h_n\partial^\lambda h_n \right.
\nn\\
&& - {1 \over 2}\left(m^2\Delta h_{n\mu\nu}\Delta h_n^{\mu\nu} -
\left(\Delta h_n\right)^2 \right) - 2 \left(m\Delta^\dagger A_n^\mu
+
\partial^\mu \varphi_n\right) \left(\partial^\nu h_{n\mu\nu} -
\partial_\mu h_n\right)
\nn\\
&& \left. - {1 \over 2}\left(\partial_\mu A_{n\nu} - \partial_\nu
A_{n\mu}\right) \left(\partial^\mu A_n^\nu - \partial^\nu
A_n^\mu\right) \right]\ . \eeq The $\Delta$ and $\Delta^\dagger$ are
difference operators, which operate on the indices $n$ as \beq
\label{KS1} \Delta\phi_n \equiv \sum_{k=0}^{N-1}a_k \phi_{n+k}\,,
\hs\hs \Delta^\dagger\phi_n \equiv \sum_{k=0}^{N-1}a_k \phi_{n-k}\,,
\hs\hs\sum_{k=0}^{N-1}a_k = 0\,, \eeq where the $a_k$ are $N$
constants and  the $N$ variables $\phi_n$ can be identified with
periodic fields on a lattice with $N$ sites if the  {\it periodic
boundary conditions} $\phi_{n+N}=\phi_n$ are imposed. The latter
condition in (\ref{KS1}) assures that $\Delta$ becomes the usual
differentiation operator in a properly defined continuum limit.

The eigenvalues and eigenvectors for $\Delta$ are given
by\footnote{Please note that here we use a different notation with
respect to one used in Refs.~\cite{multig} and \cite{SOD}. In fact,
in order to avoid confusion with masses, we have replaced the
eigenvalue $m$ with $\mu$ and the index $M$ with $p$.}
\beq
\label{KS3} && \Delta \phi_n^p = i\mu_p\phi_n^p\,,
\hs\hs i\mu_p=\sum_{n=0}^{N-1}a_n\,e^{2\pi inp/N}\,,\\
&&\phi_n^p=\frac{e^{2\pi inp/N}}{\sqrt{N}}\,.
\hs\hs p=0,1,2,3,\ldots
\eeq
By using (\ref{KS1}) in the latter equation and assuming
$a_n$ to be real one gets the relations
\beq
\mu_0=0\,,\hs\hs\mu_p=-\mu_{N-p}\,,\hs\hs
\mu_{N-p}=-\mu_p^*\,,
 \eeq
which, for any fixed $N$, permit to obtain the whole spectrum of the
theory.

Then we see that the Lagrangian (\ref{KS9}) describes a massless graviton
and $N-1$ massive gravitons, with masses $M_p=|\mu_p|$ ($p=1,2,\ldots, N-1$).
It must be pointed out that the massive gravitons always appear in
pairs which
share a common mass and, moreover, the complex mass parameter $\mu_p$
can be arbitrarily chosen, just by properly selecting
the coefficients $a_k$ in (\ref{KS3}) \cite{multig}.
As discussed in \cite{KS}, the multigraviton model can be regarded
as corresponding to a Kaluza-Klein theory where the extra dimension
lives in a lattice.

As a specific example, we now consider the two-brane Randall-Sundrum (RS)
model \cite{RS1}
(for a recent review see \cite{maar1}). In this model, the masses of
the Kaluza-Klein modes are given by
\beq
\label{KS16}
M_p={\pi p\over z_c}\ ,
\hs z_c=l\left(e^{\pi r_c/l} - 1 \right)\,,\hs p=0,1,2,...
\eeq
where $l$ is the length parameter of the five-dimensional AdS space and
$\pi r_c$ the geodesic distance between the two branes.

Motivated by this last equation (\ref{KS16}), we consider an
$N=2N'+1$ graviton model, with the graviton masses being given by
\beq \label{RS17} \mu_p=\left\{\begin{array}{ll} {\pi p \over z_c},\
& p=0,1,\cdots , N',
\nn\\
-{\pi(N-p)\over z_c},\ & p=N'+1, N'+2, \cdots , N-1 =2N'.
\end{array}\right.
\eeq
Those are solutions of Eq.~(\ref{KS3}), with the choice $a_0=0$
and, for any $n\geq1$,
\begin{eqnarray}
a_n &=& - {2\pi\over (2N'+1)z_c}
\Im \left\{{ \left(1 - e^{-i{2\pi N' n \over 2N'+1}}\right)e^{-i{2\pi n
\over 2N'+1}} \over 1 - e^{-i{2\pi n \over 2N'+1}}}\right\} \nn\\
&=& -{(-1)^n\:2\pi\over N z_c}\:
\frac{\sin^2\left({\pi nN'\over N}\right)}
{\sin\left({\pi n\over N}\right)}\:.
\label{RS18}\end{eqnarray}
We see that $N$ plays here the role of a cutoff of the Kaluza-Klein modes.

In previous models of deconstruction \cite{AH,KS},  mainly nearest
neighbor couplings  between the sites of the lattice have been
considered. As a consequence, on imposing a periodic boundary
condition, the lattice then looks as a circle. Departing from this
standard situation, in the model considered here we have  introduced
{\it non-nearest-neighbor} couplings among the sites. That is, a
site links to a number of other ones in a rather complicated way. In
this sense, the lattice in the present model is no more a simple
circle but it looks more like, say, a mesh or a net. Let us assume
that the sites on the lattice would correspond to points in a brane.
If the codimension of the spacetime is one, the brane should be
ordered, resembling the sheets of a book. One brane can  only
interact (directly) with the two neighboring branes. However, if the
spacetime is more complicated and/or the codimension is two or more,
the brane can directly interact with three or more branes, an
interaction that will be perfectly described by our model
corresponding to this case. For example, a site on a tetrahedron
connects directly with three neighboring sites. In this way, the
non-nearest-neighbor couplings we here consider may quite adequately
reflect the structure of the extra dimension. In this respect our
model is very general and opens a number of interesting
possibilities.

\subsection{The supergravity case}

By using the same sort of techniques described above,
the multi-graviton model can be generalized
to the supergravity case, just by starting with a supergravity theory
in 5-dimensions and implementing  deconstruction by way of replacing
the fifth spacelike dimension with a one-dimensional lattice containing
$N$-points.   A multi-supergravity model of this kind has been
proposed in Ref.~\cite{SOD}, to which  the interested reader is addressed
for details. Here we shall only write down the essential aspects which will
be used in what follows.

In the 5-dimensional linearized supergravity theory, the number of
bosonic degrees of freedom is 8, 5 due to the massless graviton and
3 due to the massless vector (gauge) field and the number of
fermionic degrees of freedom is 8 too, due to the the complex
Rarita-Schwinger field ($4 \times 2$).

The deconstruction process now consists in replacing the fifth
dimension in the action of spin two+vector+Rarita-Schwinger fields
with $N-$points and the derivatives with respect to the
corresponding variable with the operator $\lap$ as in
Eq.~(\ref{KS3}). In this way one gets a complicated action in 4
dimensions, similar to the one in Eq.~(\ref{KS9}), but with vector
and fermion parts too. It contains a spin-2 field $h_{\mu\nu}$ (the
graviton), but also scalar, vector and fermionic fields. More
precisely, in the massless sector one has 8 degrees of freedom due
to bosons (graviton (2 d.o.f.), gauge and St{\"u}ckelberg vectors
(2+2 d.o.f.), a St{\"u}ckelberg scalar and the fifth component of
the gauge field (1+1 d.o.f.) and 8 degrees of freedom due to
fermions (complex Dirac and Rarita-Schwinger fields), while in the
massive sector one has again 8+8 degrees of freedom, but only due to
a massive graviton, vector and Rarita-Schwinger fields. As in the
pure-gravity case, one has $N$ copies of such fields and their
masses ---obtained by imposing periodic boundary conditions--- are
always given by means of Eq.~(\ref{KS3}), that is \beq \label{KS3-2}
&& \Delta\phi_n^p = i\mu_p\phi_n^p\,,
\hs\hs i\mu_p=\sum_{n=0}^{N-1}a_n\,e^{2\pi inp/N}\,,\\
&&\phi_n^p=\phi_{n+N}^p=\frac{e^{2\pi inp/N}}{\sqrt{N}}\,.
\hs\hs p=0,1,2,3,\ldots
\eeq
On the other hand, for fermion fields
anti-periodic boundary conditions could also be considered.
In such case one gets a different spectrum, given by means
of the following equations
\beq
\label{KS3-3}
&& \Delta\tilde\phi_n^p =i\tilde\mu_p\phi_n^p\,,
\hs\hs i\tilde\mu_p=\sum_{n=0}^{N-1}a_n\,e^{2\pi in(p+1/2)/N}\,,\\
&&\tilde\phi_n^p=-\tilde\phi_{n+N}^p=\frac{e^{2\pi in(p+1/2)/N}}{\sqrt{N}}\,.
\hs\hs p=0,1,2,3,\ldots
\eeq
It has to be noted that with boundary conditions of this sort, there
are no massless fermions and this is a consequence of the explicitly
breakdown of global supersymmetry.

\section{The induced cosmological constant}

We now turn to the evaluation of the induced cosmological constant
for the $N-$ graviton and super-graviton models discussed in the
previous section. To this aim ---the main one in the present
paper--- we shall compute the one-loop effective potential using
zeta-function regularization \cite{zbs,report}; needless to say,
other regularization schemes could work as well. First of all, we
compute the effective potential for a free scalar field with mass
$M$, since this corresponds to the contribution of each degree of
freedom to the one-loop effective potential of our theories.

In the zeta-function regularization method, the one-loop
contribution to the effective potential is given by \beq
V^{(1)}_{eff}=-\frac1{2V}\zeta'(0|L/\mu^2)
=-\frac1{2V}\zeta'(0|L)-\frac1{2V}\,\zeta(0|L)\log\mu^2\:,
\label{v1a}\eeq $V$ being the volume of the manifold and
$\zeta(s|L)$ the zeta function corresponding to the Laplacian-like
operator $L=-\lap ^{2}+M^2$, with $M$ a positive constant. The
arbitrary parameter $\mu$ has to be introduced for dimensional
reasons. It will be fixed by renormalization at the end of the
process.

The manifold we are considering in the present paper is a flat one
with non trivial topology of the kind $\M=\R\X T^3$.
The simplest case $\M=\R^4$ has been already considered in \cite{multig,SOD}.

The  operator $L$ has the form
\beq
L=-\frac{d}{d\tau^2}+L_3\,,\hs\hs L_3=-\lap_3+M^2,
 \eeq
$\lap_3$ being the Laplace operator on $T^3$.
The zeta-function is expressed
in terms of the heat trace  via the Mellin representation.
The heat traces read
\beq
\Tr e^{-tL}=V{\cal K}(t|L)\,,\hs\hs
\Tr e^{-tL_3}=V_3{\cal K}(t|L_3)\,,\hs\hs
{\cal K}(t|L)=\frac{{\cal K}(t|L_3)}{\sqrt{4\pi t}}.
 \eeq
As a result
\begin{eqnarray}
\zeta(s|L)&=&\frac{1}{\Gamma(s)}\int_0^\infty
\,dt\:t^{s-1}\Tr e^{-tL}=
\frac{V}{\sqrt{4\pi}\Gamma(s)}\int_0^\infty
\,dt\:t^{s-3/2}{\cal K}(t|L_3)
\nn\\
&=&\frac{V\Ga(s-1/2)}{\sqrt{4\pi}\Gamma(s)}\, \tilde\ze(s-1/2|L_3)\,,
\label{zAd}\end{eqnarray}
$\tilde\ze(s-1/2|L_3)$ being the zeta-function density on $T^3$
and $V_3=(2\pi r)^3$ the ``volume'' of the torus with ``radius'' $r$.
The heat kernel and zeta function on  $T^3$ are well known. In the Appendix
\ref{S:torus}, for the reader's convenience,  we summarize
some useful representations that will be used in what follows
(for a review, see \cite{report}).

Using expressions (\ref{zAd}) and (\ref{KT3})
one realizes that the zeta function
can be  written as the sum of two terms, that is
\beq
\ze(s|L)=\ze_0(s|L)+\ze_T(s|L)\,,
 \eeq
where $\ze_0$ is the same one has on $\R^4$, namely
\beq
\ze_0(s|L)=
\frac{V\Gamma(s-2)M^{4-2s}}{16\pi^2\Ga(s)}
=\frac{VM^{4-2s}}{16\pi^2(s-1)(s-2)}\,,
\label{z0s}\eeq
while $\ze_T$ represents the contribution
due to the non-trivial topology, which  explicitly depends on
the topological parameter $r$.
Expression (\ref{z0s}) is also the leading contribution to the whole
zeta function in a power series expansion for large values of $M$.

Recalling now (\ref{toro}), we obtain
\beq
\ze_T(s|L)=\frac{V\Ga(s-3/2)\cos\pi s\,M^{4-2s}}{8\pi^{5/2}\Ga(s)}\,
\int_{1}^{\ii}\,du\;G(Mru)\,(u^2-1)^{3/2-s}\,.
 \eeq
Observe that the topological contribution vanishes at $s=0$,
and this means that
\beq
\ze(0|L)=\ze_0(0|L)=\frac{VM^{4}}{32\pi^2}\,.
 \eeq
Using (\ref{v1a}), for the one-loop effective potential we finally have
\beq
V^{(1)}_{eff}=\frac{M^4}{64\pi^2}\,\at\log\frac{M^2}{\mu^2}-\frac32\ct
-\frac{M^{4}}{12\pi^2}\,
\int_{1}^{\ii}\,du\;G(Mru)\,(u^2-1)^{3/2}\,.
\label{v1}\eeq
It is interesting to note that for scalar fields, in the large mass case
the topological contribution is always negative, and it is negligible
with respect to the standard Coleman-Weinberg term.

As we have anticipated above, the parameter $\mu$ has to be fixed
by a renormalization condition. To this aim,
here we follow Ref.~\cite{cher1}. The total one-loop effective potential
is of the form
\beq
V_{eff}=V_R(\mu)+V^{(1)}_{eff}(\mu)\,,
 \eeq
$V_R(\mu)$ being the renormalized vacuum energy. For physical reasons,
the last expression has to be independent of $\mu$, and this means that
\beq
\mu\:\frac{dV_{eff}}{d\mu}=0\:,
\eeq
from which it follows that
\beq
V_R(\mu)=V_R(\mu_R)+\frac{M^4}{64\pi^2}\,\log\frac{\mu_R^2}{\mu^2}\,,
\label{vr1a}\eeq
$\mu_R$ being the renormalization point which has to be fixed by the condition
$V_R(\mu_R)=0$. In this way, we finally get
\beq
V_{eff}=\frac{M^4}{64\pi^2}\,\at\log\frac{M^2}{\mu_R^2}-\frac32\ct
+V_T(r)\,,
\label{V4}\eeq
\beq
V_T(r)=-\frac{M^{4}}{12\pi^2}\,
\int_{1}^{\ii}\,du\;G(Mru)\,(u^2-1)^{3/2}=
-\frac{M^{2}}{16\pi^4r^2}\,\sum_{n\in\Z^3;n\neq0}\,
\frac{K_2(2\pi Mr|\vec n|)}{n^2}\,.
\label{VT}\eeq
$V_T(r)$ represents the contribution coming from the non-trivial topology,
which for scalar fields is always negative. We also note that, as a function
of the topological parameter $r$, $V_T(r)$ can reach, in principle, any
negative value.
In Eq.~(\ref{VT}), $K_\nu$ are the MacDonald's (or modified Bessel)
functions.

Before proceeding with the computation of the induced cosmological
constant corresponding to the models we have discussed in Sect. 2,
we first analyse here the behavior of $V_T(r)$ as a function of $r$.
To this aim, we  consider the two different regimes $Mr\ll1$ and
$Mr\gg1$.

For the case $Mr\ll1$, using (\ref{G0x}) in (\ref{VT}) we get \beq
V_T(r)&=&-\frac{M^4}{12\pi^2}\,\int_{1}^{\ii}\,du\,G(Mru)(u^2-1)^{3/2}
\nn\\ &=&
-\frac{1}{12\pi^2r^4}\,\int_{Mr}^{\ii}\,dx\,G(x)(x^2-M^2r^2)^{3/2}
\nn \\\hs &=& -\frac{1}{12\pi^2r^4}\,\aq
\int_{Mr}^{1}\,dx\,\at1-\frac{1}{\pi^2x^3}\ct\,\at
x^2-M^2r^2\ct^{3/2} \cp\nn\\&&\ap\hs\hs\hs
+\int_{0}^{1}\,dx\,G_0(x)x^3+\int_{1}^{\ii}\,dx\,G(x)x^3
+O(M^2r^2)\cq \nn\\&=& -\frac{1}{32\pi^2r^4}\,\aq
\sum_{n\in\Z^3;|\vec n|\neq0}\,\frac{1}{\pi^4|\vec n|^4}
+O(M^2r^2)\cq\sim -\frac{1}{64\pi^2r^4}+O(M^2/r^2)\,.
\label{small}\eeq We thus see that in this limit the leading term
does not depend on $M$, and that it can be arbitrarily large, with a
suitable choice of the parameter $r$. The series in the latter
equation has been computed numerically.

On the contrary, in the opposite regime, $Mr\gg 1$, using (\ref{VT})
and the asymptotic expansion for the Bessel function, we obtain \beq
V_T(r)&=&-\frac{M^{2}}{16\pi^4r^2}\,\sum_{n\in\Z^3;n\neq0}\,
\frac{K_2(2\pi Mr|\vec n|)}{n^2} \nn \\ \hs &\sim&
-\frac{3M^4}{32\pi^4(M r)^{5/2}}\,\,\,e^{-2\pi Mr}+ \cdots
\label{large} \eeq In this limit the topological contribution could
indeed be arbitrarily small. In Fig. 1 the whole behavior of the
topological part of the effective potential is drawn. In order to
work with dimensionless variables we have introduced the function
$\tilde{V}_T(y) \equiv r^4V_T(r)$ of the dimensionless variable
$y\equiv Mr$. The graphic corresponds to the exact expression for
$\tilde{V}_T(y)$, as given e.g. by the first lines of Eq.
(\ref{small}), multiplied by $3\cdot 64 \pi^2$. A very smooth
transition from the behavior corresponding to $Mr\ll1$, Eq.
(\ref{small}), to the one for $Mr\gg1$, Eq. (\ref{large}), is
revealed.
\begin{figure}[htb]
\centerline{\epsfxsize=15cm \epsfbox{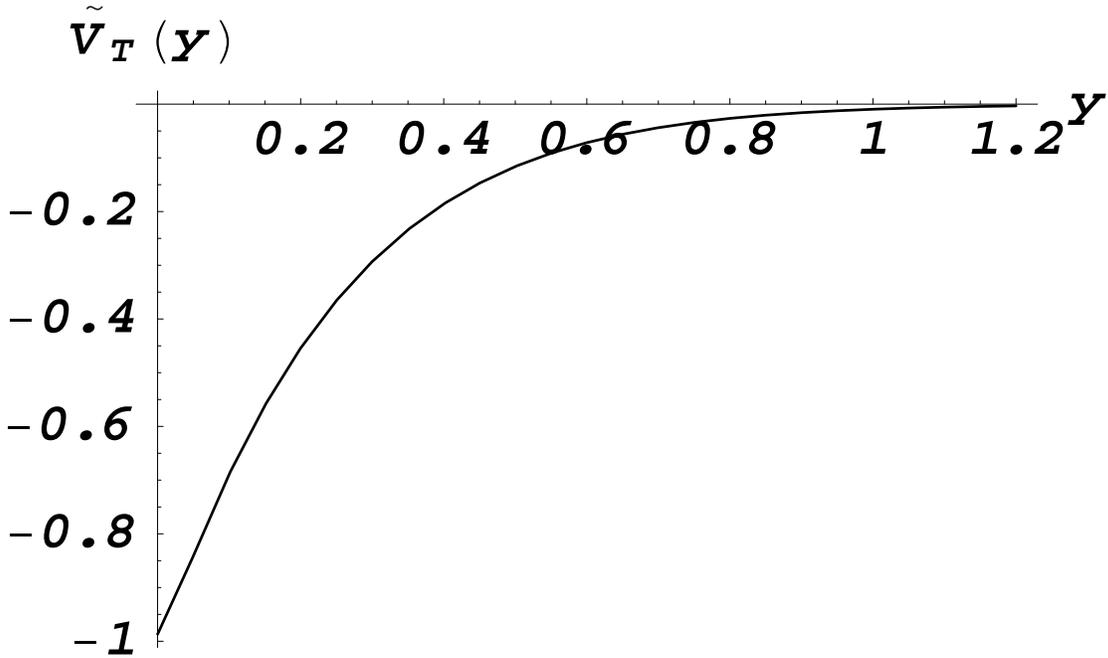}}
\caption{{\protect\small The exact expression for $\tilde{V}_T(y)\equiv
r^4V_T(r)$, multiplied by $3\cdot64 \pi^2$, as a function of $y\equiv
Mr$. }} \label{f1}
\end{figure}
In Fig. 2, the corresponding graphic of the full effective
potential, Eq. (\ref{V4}), is drawn, again as a function of $y$ and
setting $\mu_R r=1$.
\begin{figure}[htb]
\centerline{\epsfxsize=15cm \epsfbox{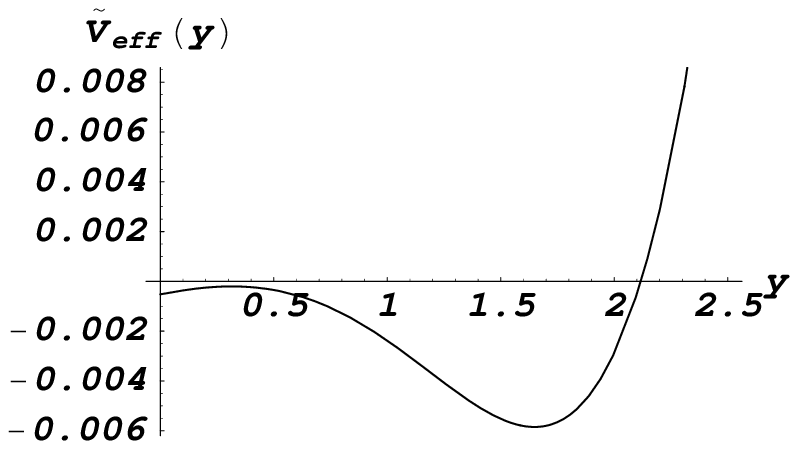}}
\caption{{\protect\small The exact expression for
$\tilde{V}_{eff}(y)\equiv
r^4V_{eff}(r)$, Eq. (\ref{V4}), as a function of $y\equiv Mr$. }}
\label{f2}
\end{figure}

At this point, the effective potential $-$and, as a consequence, the
induced cosmological constant for the models we are interested in$-$
can be obtained by adding up several contributions of the kind
(\ref{V4}).

\subsection{The multi-graviton model} We start with the explicit
example of multi-graviton model given by  (\ref{RS17}), in which
there is a single
 massless graviton and $(N-1)/2$ couples of massive gravitons,
with masses given by
\beq
M_0=|\mu_0|=0\,,\hs\hs
M_p=|\mu_p|=\frac{\pi p}{z_c}\,,\hs\hs p=1,2,...,\frac{N-1}{2}\,.
 \eeq
On the manifold $\M=\R\X T^3$, the massless graviton  gives no
contribution to the effective potential, while it does appear
explicitly on manifolds with a non-vanishing curvature.
Since the massive gravitons always show up in pairs,
in order to perform the computation of the effective potential,
it is sufficient to consider only one half of the whole massive spectrum.
Moreover, we have to take into account that each massive graviton
contributes with five scalar degrees of freedom.
After these considerations have been properly taken into account,
for the effective potential we get the following expression
\beq
V_{eff}&=&10\sum_{p=1}^{\frac{N-1}2}\frac{M^4_p}{64\pi^2}
\left(\ln\frac{M_p^2}{\mu_R^2}-\frac{3}{2}\right)\nn\\
&&\hs\hs -10\sum_{p=1}^{\frac{N-1}2}\frac{M_p^{4}}{12\pi^2}\,
\int_{1}^{\ii}\,du\;G(M_p ru)\,(u^2-1)^{3/2}\,.
\label{veff1}\eeq
One can see that, as for the non-compact flat case
(see Ref.~\cite{multig} for details), in order to have a
(small) positive induced cosmological constant one has
to consider a large value of $N$, that is, a huge number of massive
gravitons. In this respect, the torus topology does not improve
the situation. As we are now going to show, this is no longer
 the case for the multi-supergraviton model.

\subsection{The multi-supergraviton model} Here we have to
distinguish two cases: the first one corresponds to the choice of
periodic boundary conditions in both the bosonic and fermionic
sectors. In such situation, the degrees of freedom due to bosons
exactly compensate the degrees of freedom due to fermions. Moreover,
for any massive boson there is a fermion with the same mass and,
since the contribution to the effective potential  of any fermionic
degree of freedom is opposite to the contribution of a bosonic
degree of freedom, it turns out that the induced cosmological
constant vanishes, independently of the mass spectrum.

In the second case, that is when anti-periodic boundary conditions are
imposed in the fermionic sector, the situation changes completely,
since the fermionic mass spectrum becomes really different
with respect to the bosonic one.
For example, by choosing $N=3$ \cite{SOD},
the solutions of Eqs.~(\ref{KS3-2}) and (\ref{KS3-3}) give
\beq
M_0=0\,,\hs\hs M_1=M_2=m\,,\hs\mbox{for bosons}\,,
\eeq
\beq
\tilde M_0=\tilde M_2=\frac{m}{\sqrt3}\,,\hs\hs
\tilde M_1=\frac{2\,m}{\sqrt3}
\,,\hs\mbox{for fermions}\,,
 \eeq
$m$ being an arbitrary mass parameter.

By taking into account the number of degrees of freedom, the
one-loop effective potential becomes, in this case \beq
V_{eff}&=&\frac{M^4_1}{4\pi^2}
\left(\ln\frac{M_1^2}{\mu_R^2}-\frac{3}{2}\right)
-\frac{4M_1^{4}}{3\pi^2}\, \int_{1}^{\ii}\,du\;G(M_1
ru)\,(u^2-1)^{3/2} \nn\\&&\hs -\frac{\tilde M^4_0}{4\pi^2}
\left(\ln\frac{\tilde M_0^2}{\mu_R^2}-\frac{3}{2}\right)
+\frac{4\tilde M_0^{4}}{3\pi^2}\, \int_{1}^{\ii}\,du\;G(\tilde M_0
ru)\,(u^2-1)^{3/2} \nn\\&&\hs\hs -\frac{\tilde M^4_1}{8\pi^2}
\left(\ln\frac{\tilde M_1^2}{\mu_R^2}-\frac{3}{2}\right)
+\frac{2\tilde M_1^{4}}{3\pi^2}\, \int_{1}^{\ii}\,du\;G(\tilde M_1
ru)\,(u^2-1)^{3/2} \nn\\&=&
-\frac{m^4}{36\pi^2}\,\log\frac{2^{16}}{3^9}+V_T\,,
\label{veff2}\eeq $V_T$ being the sum of all the topological
contributions. As one sees, the first term on the right-hand side of
the latter equation is always negative, but the whole effective
potential could be positive due to the presence of the topological
term. For example, in the regime $mr\ll1$, from (\ref{small}) one
has \beq V_T\sim\frac{1}{8\pi^2r^4}\segue V_{eff}>0\hs\mbox{for}\hs
mr<\at\frac{2}{9}\,\log\frac{2^{16}}{3^9}\ct^{-1/4}\sim 1.4\,,
 \eeq while in the opposite regime, $mr\gg1$, by using
(\ref{large}) one can see that the topological contribution although
still positive it is negligible, and thus the effective potential
remains negative.

In Fig. 3, the corresponding graphic of the full effective
potential, Eq. (\ref{veff2}), is drawn, again as a function of
$y\equiv mr$.
\begin{figure}[htb]
\centerline{\epsfxsize=15cm \epsfbox{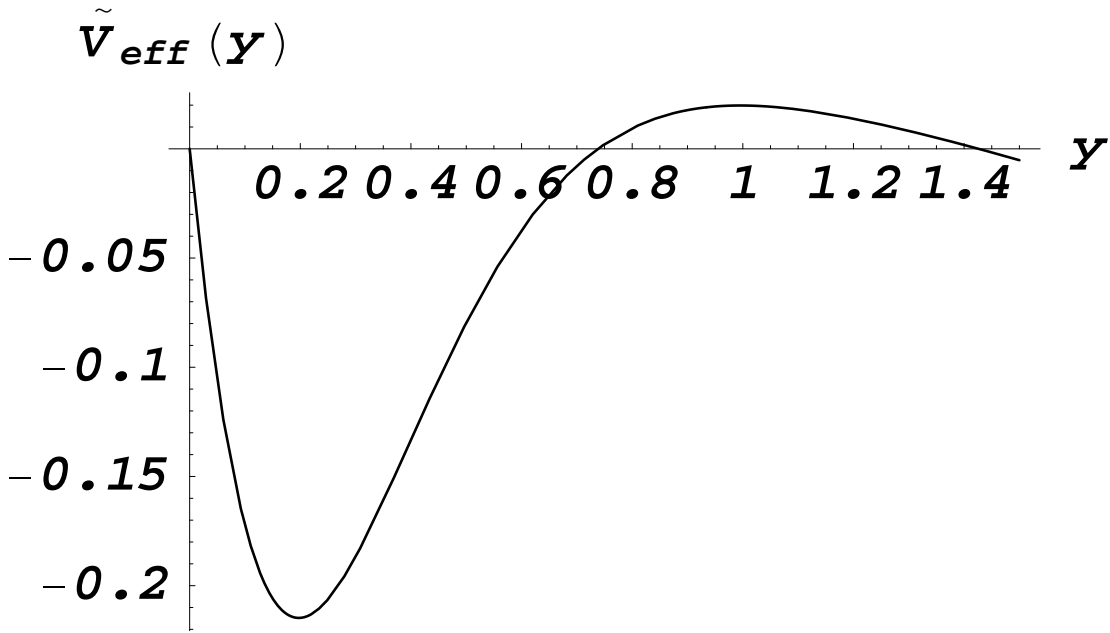}}
\caption{{\protect\small The exact expression for
$\tilde{V}_{eff}(y)\equiv
r^4V_{eff}(r)$, Eq. (\ref{veff2}), as a function of $y\equiv mr$. }}
\label{f3}
\end{figure}

\section{Conclusions}

In this paper, we have computed the effective potential corresponding to
 a multi-graviton
model with supersymmetry in the case where the bulk is a flat manifold with
the topology of a torus (more precisely $\R\X T^3$), and we have shown
that the  induced cosmological constant could be positive due to topological
contributions.

In previous papers \cite{multig,SOD} multi-graviton and
multi-supergraviton models have been considered in $\R^4$. It has
been shown that in the multi-graviton model the induced cosmological
constant can be positive, but only if the number of massive
gravitons is sufficiently large, while in the supersymmetric case
the cosmological constant can be positive
if one imposes anti-periodic boundary conditions in
the fermionic sector.
Note that the topological effects discussed above may also be relevant
in the study of electroweak symmetry breaking in models
with a similar type of non-nearest-neighbour couplings, for the
 deconstruction issue \cite{sugamoto2004}.

In the case of the torus topology, the topological contributions to
the effective potential have always a fixed sign, depending on the
boundary conditions one imposes. In fact, they are negative for
periodic fields and positive for anti-periodic fields. This means
that the torus topology provides a mechanism which, in a most natural
way, permits to have a positive cosmological constant in the
multi-supergravity model with anti-periodic fermions, being the value
of such constant regulated by the corresponding size of the
torus.\footnote[1]{A more crude analysis for the pure scalar
case already
hinted towards this conclusion. However, the sign issue was there not
easy to fix \cite{mov1}, the reason being now clear.}
In this situation one can most naturally use the minimum number, $N=3$,
of copies of bosons and fermions.

We finish  with the remark that it would be interesting to
apply the deconstruction scheme of Ref.~\cite{multig} also
for the case of two latticized extra dimensions,
which in the continuous limit would contain the orbifold
singularity. This analysis might have a quite interesting impact on
 brane running coupling calculations \cite{milton2002}.

\ack
We thank Sergei D.~Odintsov for useful discussions and suggestions.
Support from the program INFN(Italy)-CICYT(Spain), from DGICYT (Spain),
project BFM2003-00620, and from SEEU grant PR2004-0126 (EE), is gratefully
acknowledged.


\appendix
\s{Zeta function on the torus}
\label{S:torus}

Eigenvalues of the Laplacian on the 3-dimensional
torus are of the form $\la_n=n^2$, $n\in \Z^3$,
and thus the corresponding heat kernel is given by
\beq
{\cal K}(t|L_3)=\frac{e^{-tM^2}}{V_3}
\sum_{n\in\Z^3}\,e^{-tn^2/r^2}
=\frac{e^{-tM^2}}{(4\pi t)^{3/2}}\,
\sum_{n\in\Z^3}\,e^{-\pi^2n^2r^2/t}\,,
\label{KT3}
\eeq
being $V_3=(2\pi r)^3$ the ``volume'' of $T_3$.
Using the expression above, the zeta function of this Laplacian
can be written as
\beq
\ze(s|L_3)=\ze_0(s|L_3)+\ze_T(s|L_3)\,,
 \eeq
where the contribution
\beq
\ze_0(s|L_3)=\frac{V_3\Gamma(s-3/2)M^{3/2-2s}}{(4\pi)^{3/2}\Ga(s)}\,,
\label{z00s}\eeq
comes from the $n=0$ term and it is the same one has for $\R^3$,
while $\ze_T$ corresponds to the contribution
due to the non-trivial topology.
Such term can be written in different ways,
for instance, as an infinite sum of Bessel functions.

In Refs.~\cite{zbs} one can find many interesting results concerning
zeta functions and heat kernels corresponding to operators on manifolds
with constant curvature.
In particular, on the torus one has the very nice representation
\beq
\tilde\ze_T(z|L_3)&=&
\frac{M^{3-2z}\sin\pi z}{4\pi^2(1-z)}\,
\int_{1}^{\ii}\,du\;G(Mru)\,(u^2-1)^{1-z}
\nn\\&=&
-\frac{\pi^{z-2}}{4\Ga(z)}\sum_{n\in\Z^3;n\neq0}\,
\at\frac{M}{r|\vec n|}\ct^{3/2-z}\,\,K_{3/2-z}(2\pi M r|\vec n|)\,,
\label{toro}\eeq
where $G(x)$ is given by
\beq
G(x)=\sum_{n\in\Z^3;\,n\neq0} e^{-2\pi|\vec n|x}&=&
-1+\frac{x}{\pi^2}\,\sum_{n\in\Z^3}\,\frac{1}{(n^2+x^2)^2}
\nn\\&=&
-1+\frac{1}{\pi^2x^3}+\frac{x}{\pi^2}\,G_0(x)\,,
\label{G0x}
\eeq
$G_0(x)$ being the regular function
\beq
G_0(x)=\sum_{n\in\Z^3;\,n\neq0}\,\frac{1}{(n^2+x^2)^2}\,.
\eeq

\end{document}